\documentclass[11pt]{article}
\usepackage{moriond,graphicx}
\usepackage{xspace}
\usepackage{url}

\def\be{\begin{equation}}
\def\ee{\end{equation}}
\def\bea{\begin{eqnarray}}
\def\eea{\end{eqnarray}}

\newcommand{\half}{\ensuremath{\frac{1}{2}}\xspace}
\newcommand{\babar}{{\sc BaBar}\xspace}

\newcommand{\mathsl}[1]{\mbox{\textsl{#1}}}

\newcommand{\meson}[1]{\ensuremath{\mathsl{#1}}}

\newcommand{\lepton}[1]{\ensuremath{\mathsl{#1}}}

\newcommand{\pion}{\ensuremath{\pi}\xspace}
\newcommand{\piplus}{\ensuremath{\pion^+}\xspace}
\newcommand{\piminus}{\ensuremath{\pion^-}\xspace}
\newcommand{\pizero}{\ensuremath{\pion^0}\xspace}

\newcommand{\kaon}{\ensuremath{\meson{K}}\xspace}

\newcommand{\kminus}{\ensuremath{\kaon^-}\xspace}
\newcommand{\kplus}{\ensuremath{\kaon^+}\xspace}

\newcommand{\dmeson}{\ensuremath{\meson{D}}\xspace}
\newcommand{\dplus}{\ensuremath{\dmeson^{+}}\xspace}
\newcommand{\dzero}{\ensuremath{\dmeson{}^{0}}\xspace}

\newcommand{\dsp}{\ensuremath{\dmeson^{*+}}\xspace}
\newcommand{\dsplus}{\ensuremath{\dmeson_{s}^{+}}\xspace}

\newcommand{\electronneg}{\lepton{e}^-\xspace}
\newcommand{\positron}{\lepton{e}^+\xspace}

\newcommand{\kpipi}{\ensuremath{\dplus \to  \kaon^- \piplus \piplus}\xspace}

\newcommand{\kpi}{\ensuremath{\dzero \to  \kaon^- \piplus }\xspace}
\newcommand{\kpipipi}{\ensuremath{\dzero \to  \kaon^- \piplus \piplus \piminus}\xspace}
\newcommand{\photon}{\ensuremath{\gamma}\xspace}

\newcommand{\figref}[1]{Figure~\ref{fig:#1}}
\newcommand{\tabref}[1]{Table~\ref{tab:#1}}

\newcommand{\figlabel}[1]{\label{fig:#1}}
\newcommand{\tablabel}[1]{\label{tab:#1}}

\newcommand{\mevcc}{\ensuremath{\mathrm{MeV}/c^2}\xspace}

\newcommand{\gevcc}{\ensuremath{\mathrm{GeV}/c^2}\xspace}

\begin{document}
\vspace*{4cm}
\title{CHARMED HADRON SPECTROSCOPY FROM FOCUS}

\author{ E.W. VAANDERING \footnote{For the FOCUS Collaboration
(\texttt{http://www-focus.fnal.gov/})} }

\address{Department of Physics and Astronomy,\\
Vanderbilt University,\\
6301 Stevenson Center,\\
Nashville, TN 37235, U.S.A.}

\maketitle\abstracts{
We present charmed hadron spectroscopy results from the photoproduction
experiment FOCUS (FNAL-E831). We report new, precise measurements of the masses
and widths of the $\dmeson_2^{*+}$ and $\dmeson_2^{*0}$ mesons, evidence for the
 $\dmeson_0^{*+}$ and $\dmeson_0^{*0}$ broad states (the first such evidence in
 $\dzero \piplus$), and confirmation of the recently observed
 $\dmeson_s^+(2317)$ charmed-strange state. 
}

\section{Introduction}

With very accurate measurements of the parameters of the ground state charmed
mesons, interest has shifted to the array of excited charm meson states,
especially with the interesting discoveries of the last year in the $D_s$
sector. 

In the limit of infinite quark masses, $D$ mesons may be treated as a ``hydrogen
atom'' type object where the heavy quark does not contribute to the orbital
degrees of freedom. In this treatment the quantum numbers of the heavy and
light quarks decouple. The heavy quark is described by its spin, $s_Q$, and the
light quark degrees of freedom are described by {\boldmath$ j_q = s_q + L$ where
$s_q$ is the spin of the light quark and   $L$} is its
angular momentum. 
For $L=1$ we have $j_q = 1/2, 3/2$. Combined with $s_Q = 1/2$, we obtain four
states, one with $J=0$, one with $J=2$, and two with $J =1$ (one each with $j_q
= 1/2$ and $3/2$). In the Heavy Quark Symmetry limit, conservation of parity and
$j_q$ requires that the strong decays $D_J^{(*)}(j_q = 3/2) \to D^{(*)} \pi$
proceed via $D$-wave decays while   $D_J^{(*)}(j_q = 1/2) \to D^{(*)} \pi$
proceed via $S$-wave decays. States decaying via $S$-wave decays are expected to
be broad while those decaying via $D$-wave  are expected to be narrow. 

\section{Measurements of $L=1$ \dmeson mesons decaying to $D \pi$}

Photoproduction of charm is a good compromise between the excellent purity of
$\electronneg \positron$ collisions and the large numbers of higher multiplicity
events available in hadron-nucleon collisions. The lower multiplicity  of the
photoproduction vertex is especially important for spectroscopy of excited charm
states since discriminating between pions produced in the interaction and those
originating from decays is difficult.

For its studies of $\dzero \piplus$ and $\dplus \piminus$ final states, FOCUS
begins with over 500,000 \dmeson mesons decaying into their ``golden'' modes:
\kpi, \kpipi, and \kpipipi. These samples are shown in \figref{d2_d_sample}.
Typical cuts are placed on these samples requiring the production and decay
vertices to be well separated and that the daughter particle species are identified by
the \v{C}erenkov system. Combinations within $2\sigma$ of the nominal
\dmeson mass are combined with additional pions to form excited \dmeson meson
candidates. Also shown in \figref{d2_d_sample} is the invariant mass of $\dsp
\to \dzero \piplus$ candidates; \dzero{}s from candidates within  $3\sigma$
of the \dsp mass  are excluded from further consideration.

\begin{figure}
  \begin{center}
    \includegraphics[width=6cm]{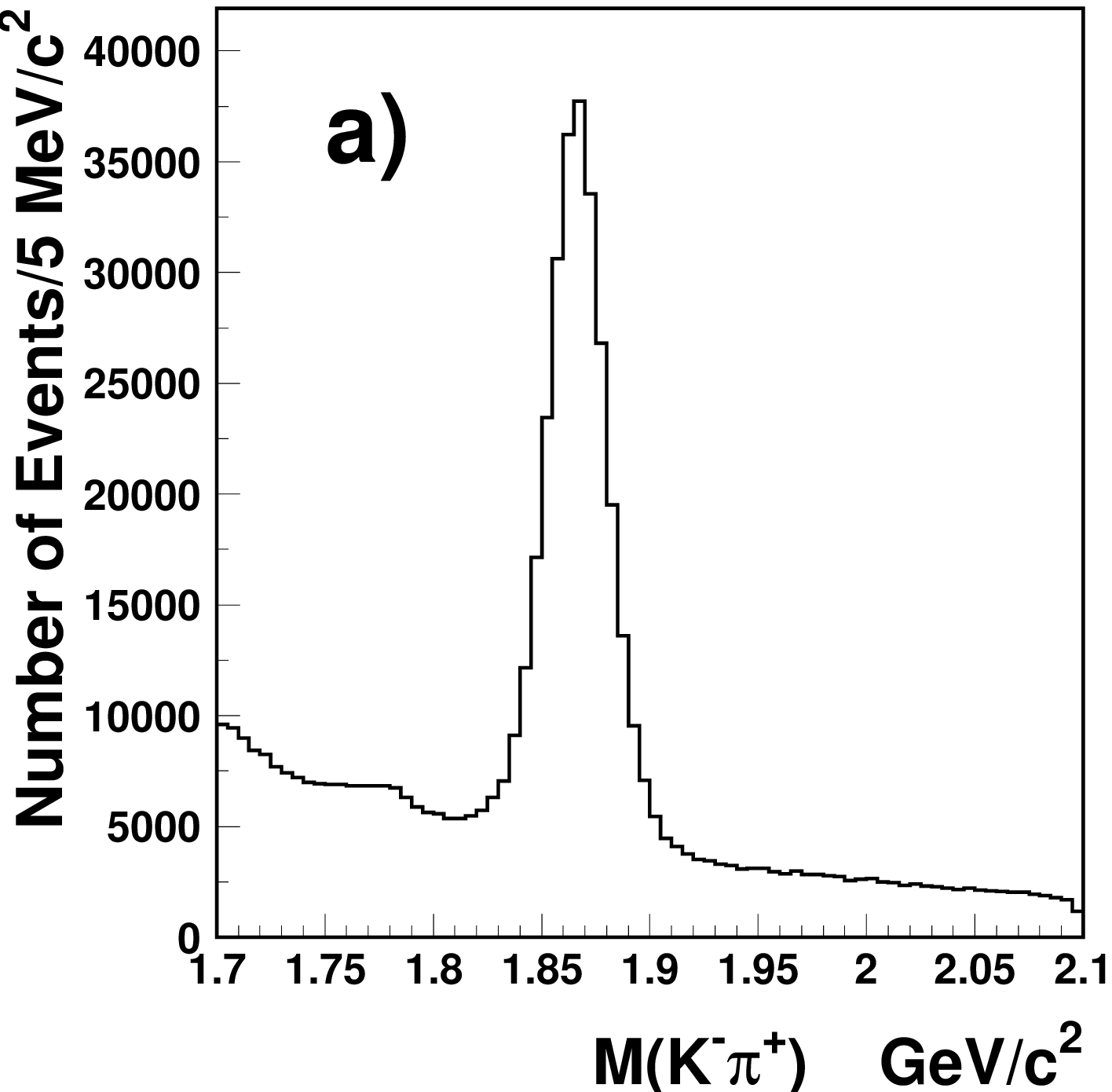}  
    \includegraphics[width=6cm]{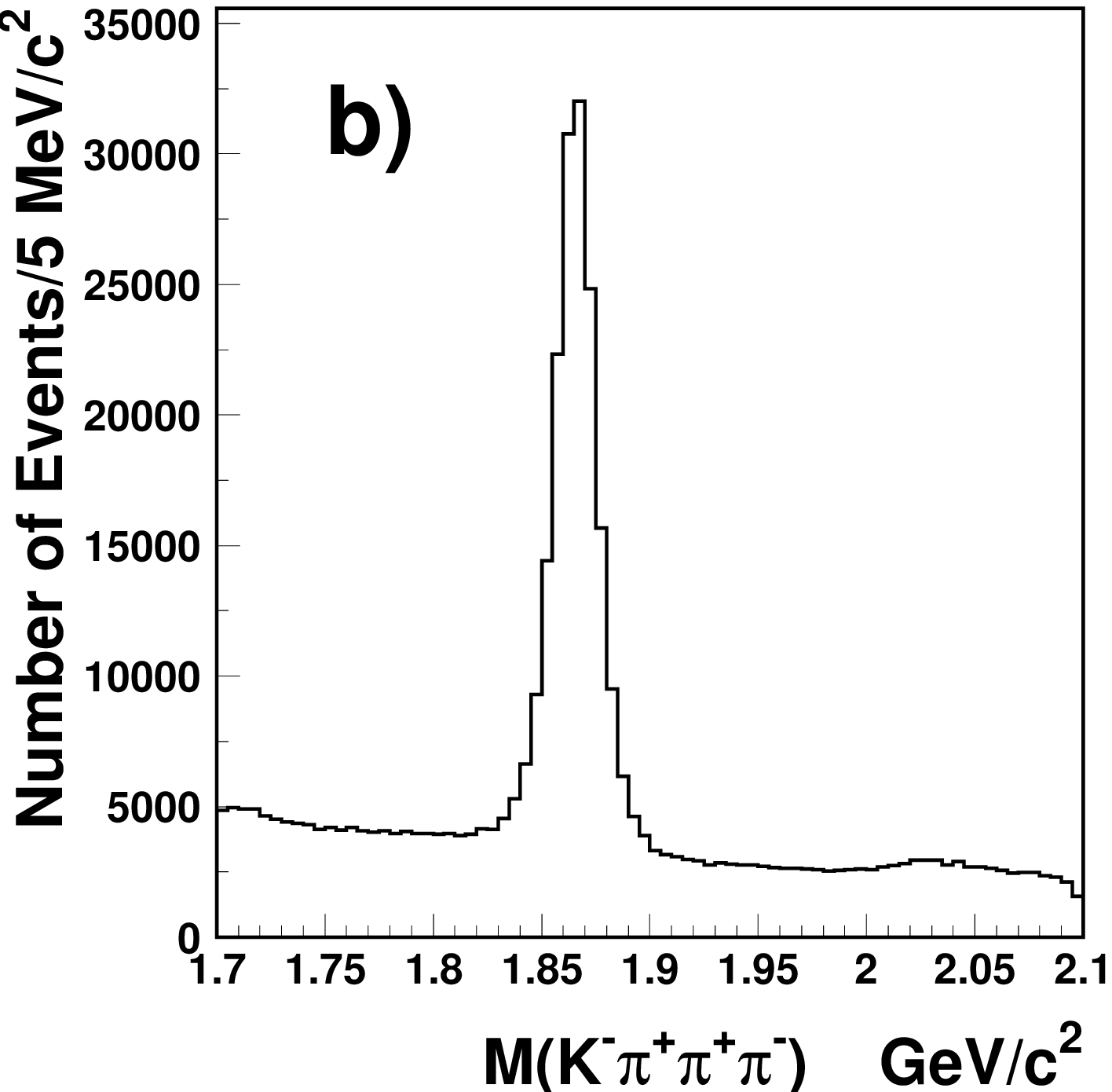}  
\\
    \includegraphics[width=6cm]{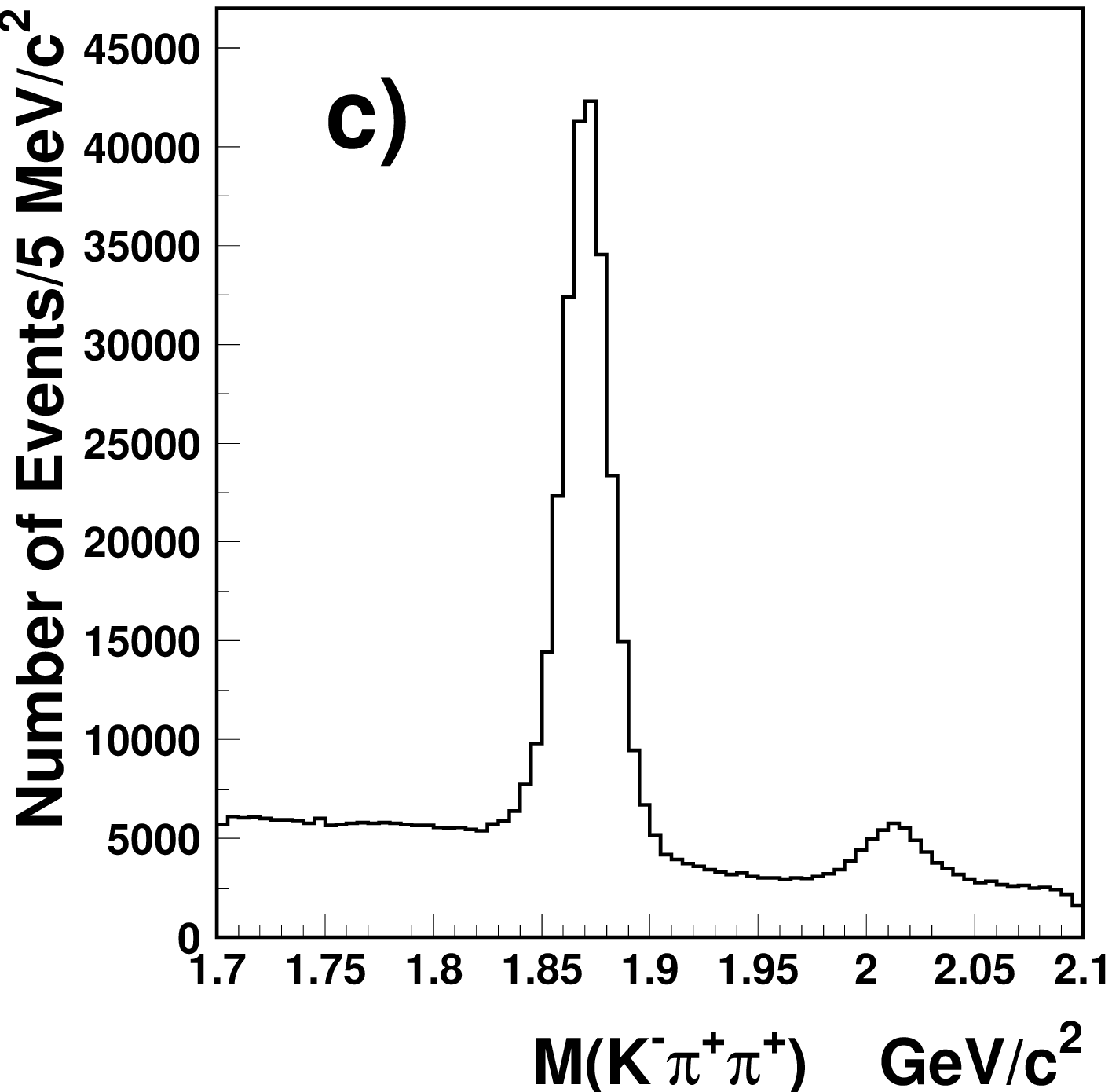}
    \includegraphics[width=6cm]{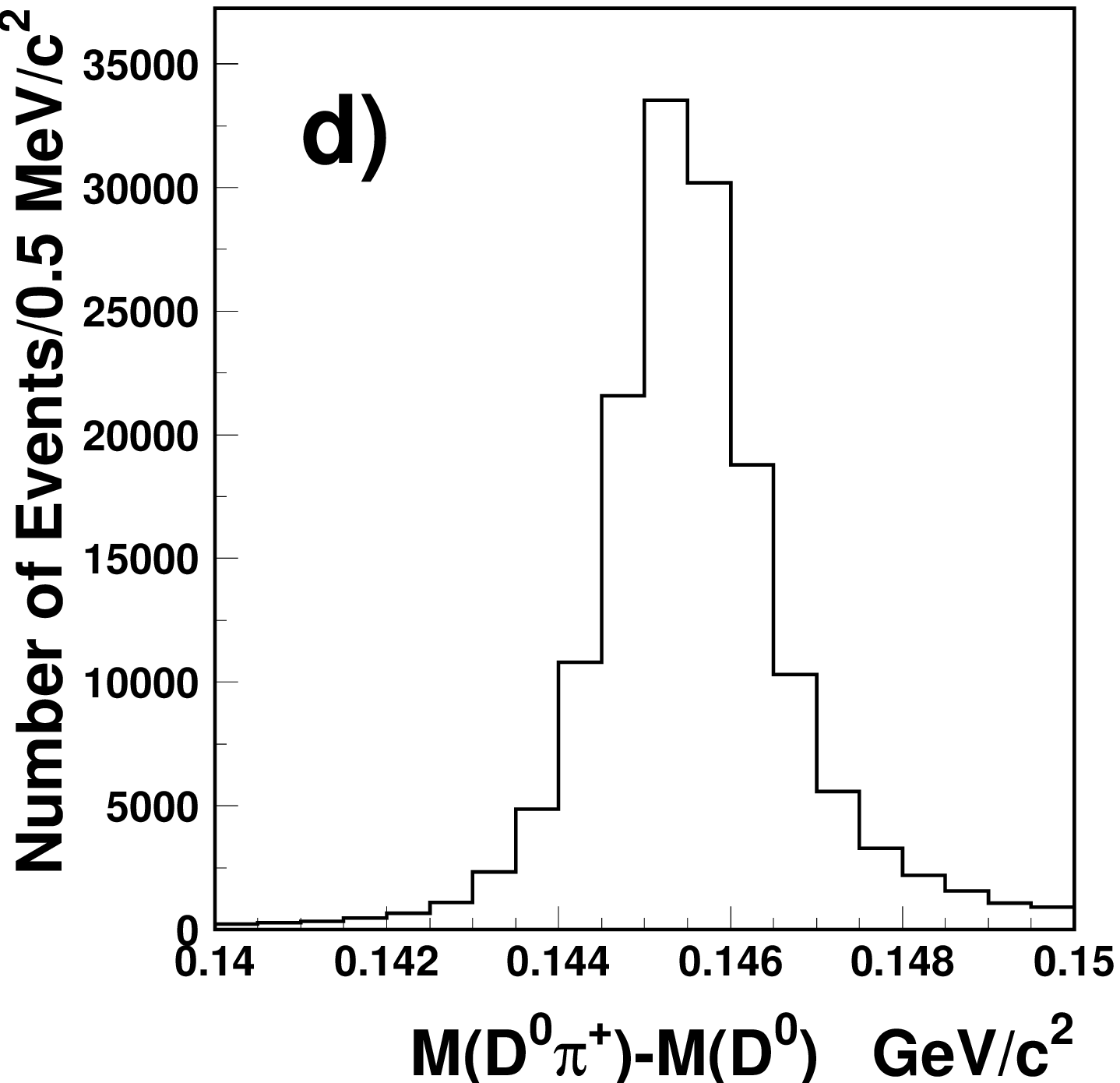}
  \end{center}  
  \caption{\dzero and \dplus samples used in the analysis. The top left plot
  shows \kpi, the top right shows \kpipipi, and the bottom left shows the
  \kpipi sample. Also shown is the $\dsp -\dzero$ mass difference which is used
  as a veto on \dzero candidates. (The additional bump in the \kpipi plot is
  from $\dsp \to \dzero \piplus$; $\kpi$.)}
  \figlabel{d2_d_sample}
\end{figure}

In \figref{d2_nosw} we show the results of fitting the $\dmeson \pion$
invariant mass distributions without including contributions from broad
$S$-wave decays. The left-most ``peak'' in both plots is due to feed-downs from
the $D_1$ and $D_2^*$ states which decay to $D^* \pi$ and the $D^*$
subsequently decays to either a \dzero or \dplus and an undetected \photon or
\pizero. The shape of this feed-down contribution is determined by Monte Carlo
simulations using the PDG values for the $D_1$ and $D_2^*$ masses and widths.
The second, right-most peaks, are the previously observed $D_2^*$ states. 
Inspecting the fits in \figref{d2_nosw}, it is apparent that the fit quality is
very poor between the feed-down and $D_2^*$ peaks for both charge combinations
even though the  $D_2^*$  fit parameters obtained in this fit are very far from
the accepted values. It is this disagreement that leads us to introduce an $S$-wave
contribution.

\begin{figure}
  \begin{center}
  \includegraphics[width=7.9cm]{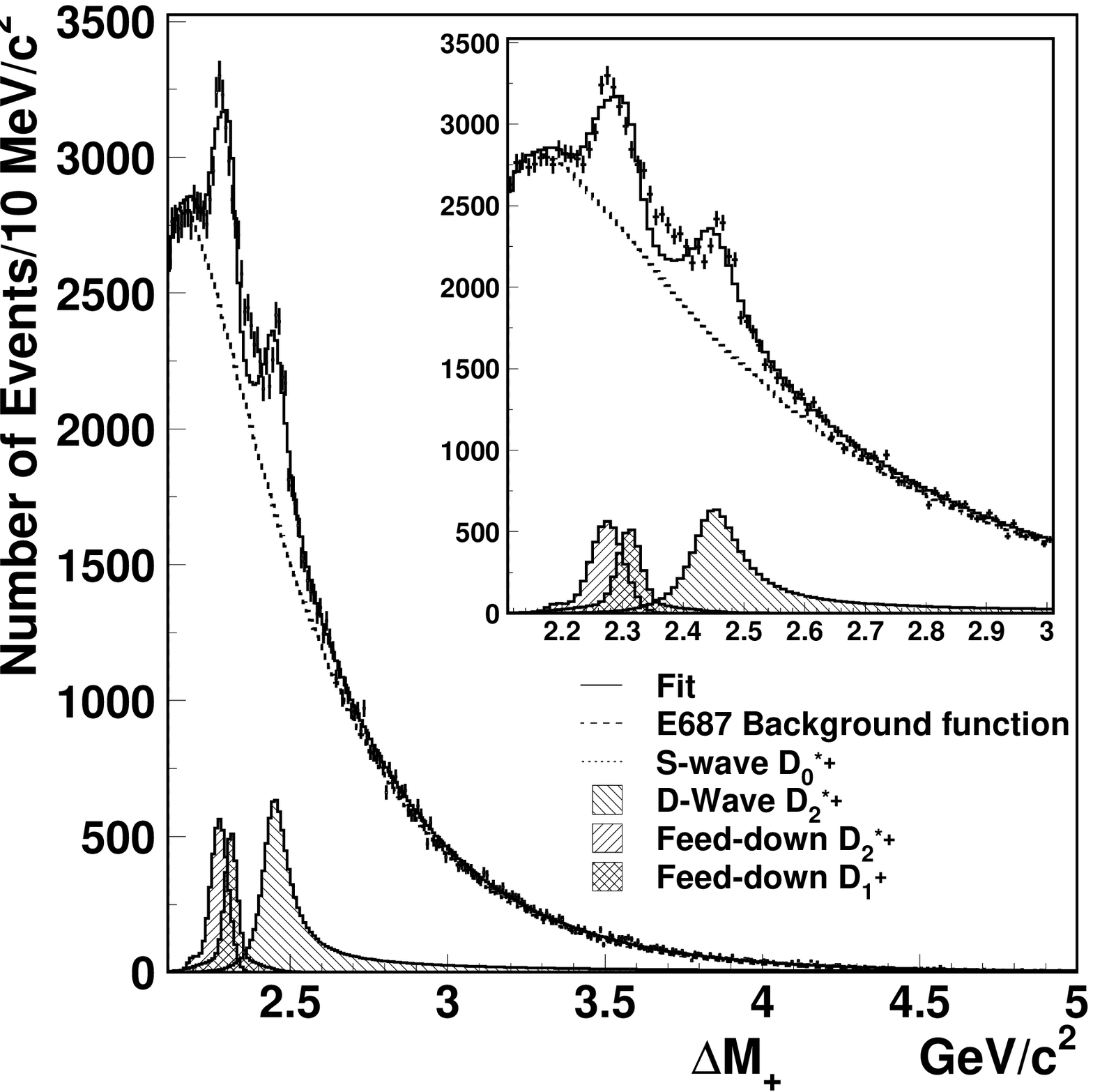}  
  \includegraphics[width=7.9cm]{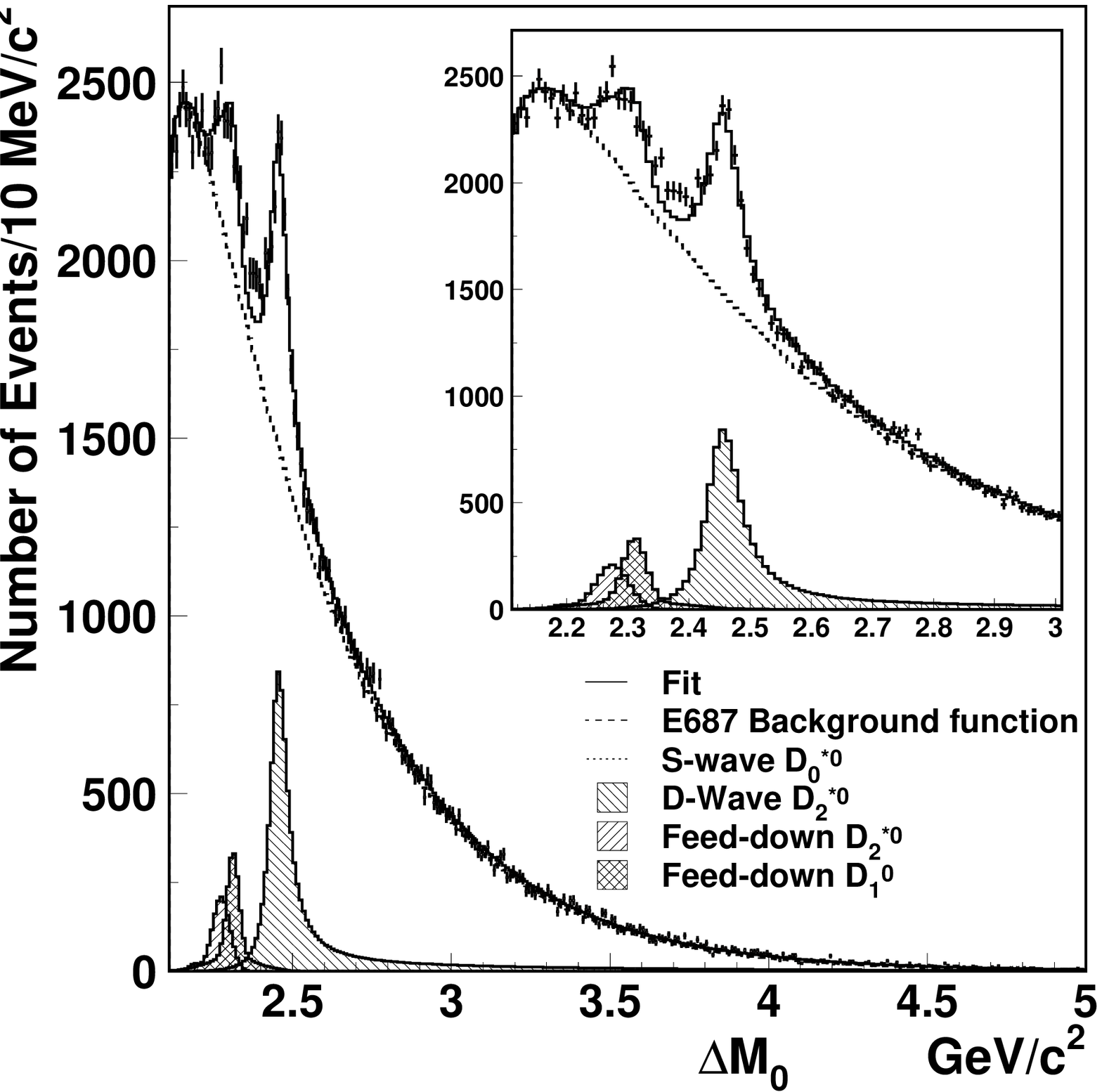}
  \end{center}  
  \caption{Fitting $\dmeson \pion$ invariant masses without an $S$-wave
  contribution. The $\dzero\piplus$ invariant mass is shown on the left, the
  $\dplus\piminus$ on the right. Also shown in the insert of each plot is the
  invariant mass below 3.0~\gevcc.}
  \figlabel{d2_nosw}
\end{figure}
  
In \figref{d2_final} we show the results obtained by fitting with the
contributions outlined above plus an additional contribution attributed to
$D_0^* \to \dmeson \pion$ decays. Not, however, that we cannot distinguish
between $D_0^* \to \dmeson \pion$ and other decays, such as a $j_q = 1/2$
(broad) $D_1 \to D^* \pion$ where the $D^*$ decays with undetected neutrals.
However, the measured yields of excess in neutral and charged final states
suggest the contamination from feed-downs is small. In our measurement, shown
in  \figref{d2_final}, we have also included our values for the $D_2^*$
parameters in the simulated feed-down shapes rather than PDG values as in
\figref{d2_nosw}. (This is a small effect.)

We have tried several different background parameterizations and other
systematic tests to assess the errors on our measurements and to test the
assumption of a broad component. In all cases, a broad component is needed  to
adequately fit the data. To minimize systematic errors on the mass scaling of
the experiment, we actually measure mass differences with respect to the \dzero
or \dplus and add the PDG \dmeson masses to obtain our final numbers. As no one
source of systematic error dominates, the reader is referred to the
publication~\cite{Link:2003bd} for a detailed discussion of the systematic
studies.

\begin{figure}
  \begin{center}
  \includegraphics[width=7.9cm]{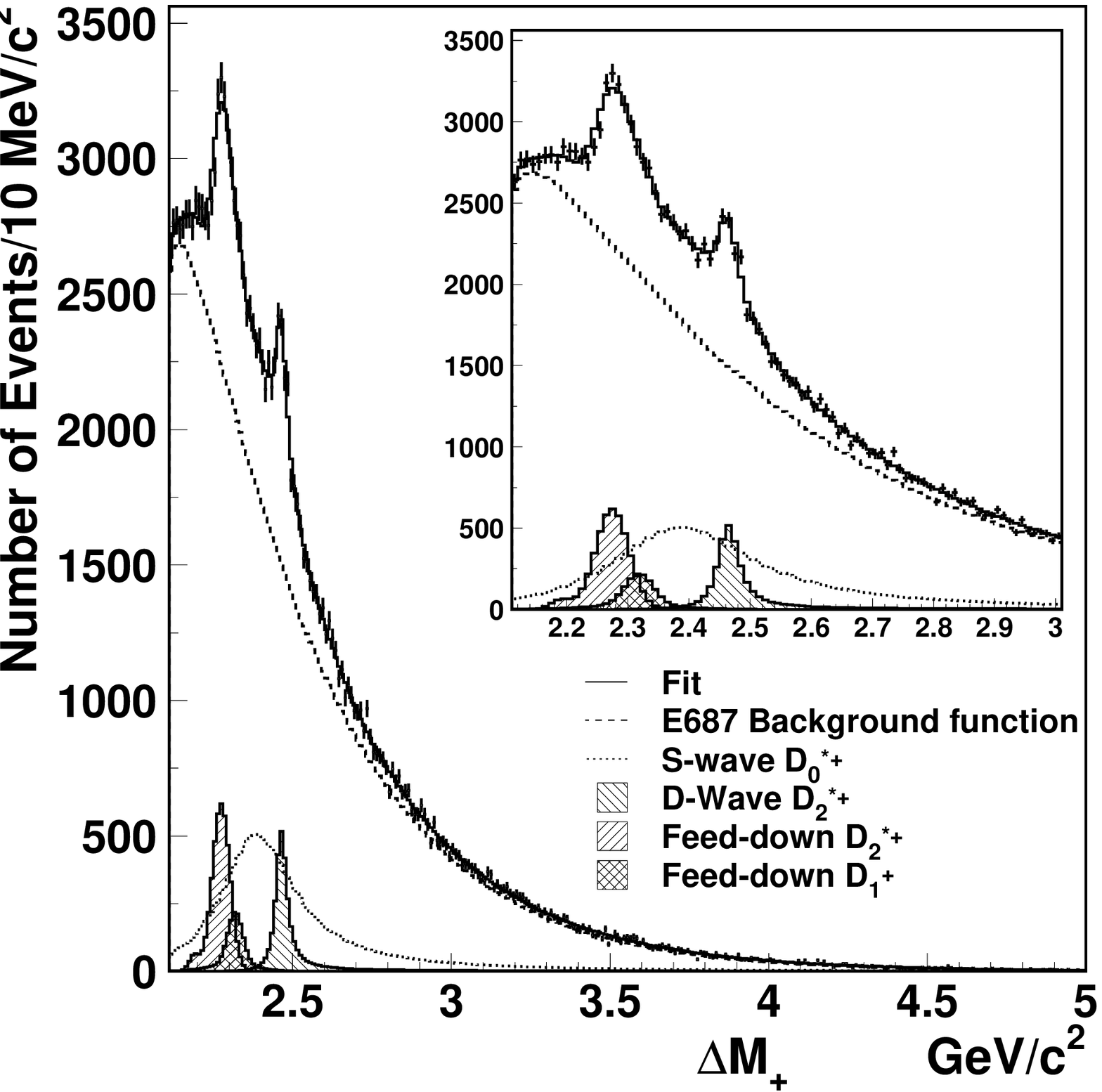}  
  \includegraphics[width=7.9cm]{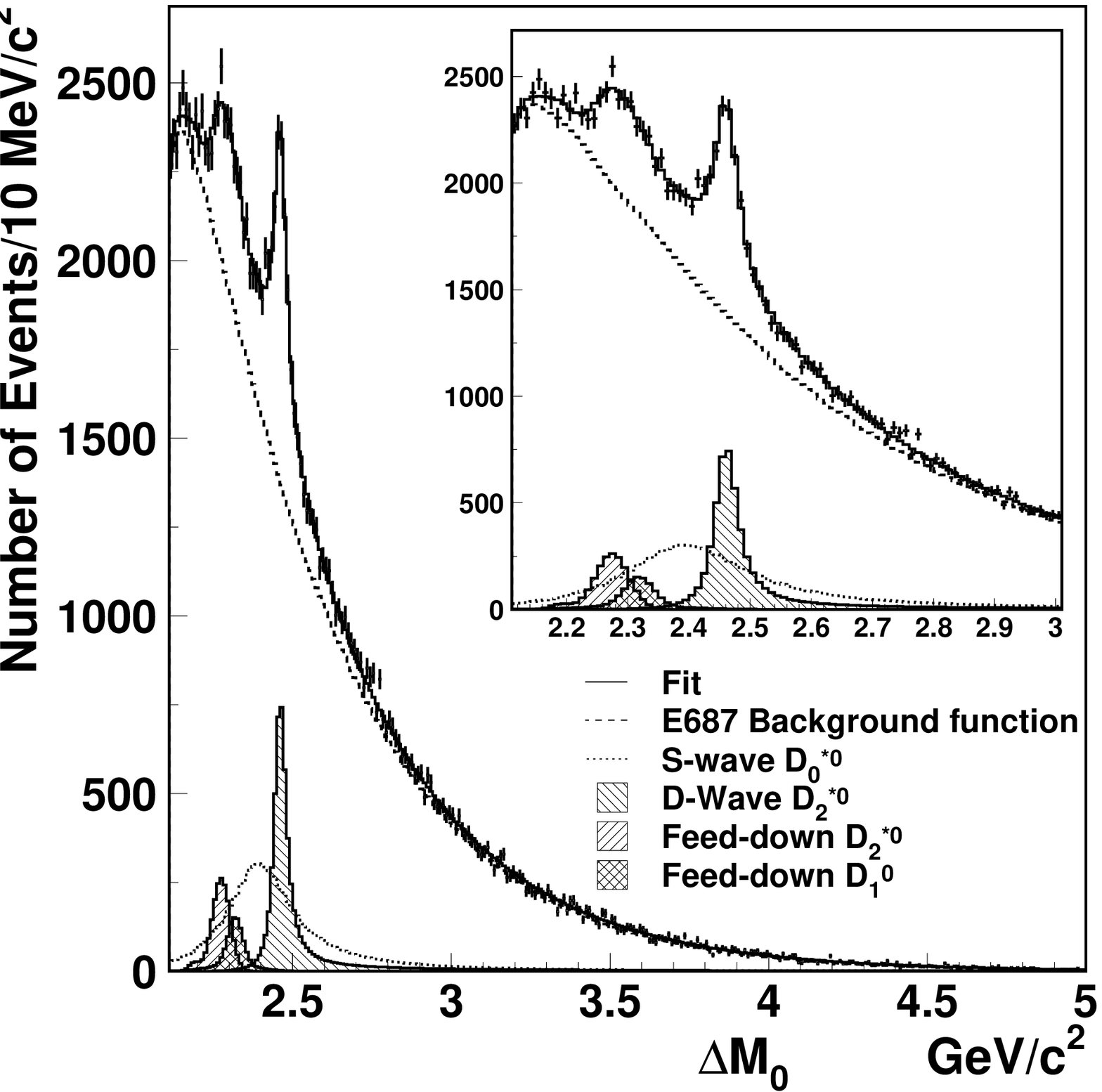}
  \end{center}  
  \caption{Fitting $\dmeson \pion$ invariant masses with an $S$-wave
  contribution. The $\dzero\piplus$ invariant mass is shown on the left, the
  $\dplus\piminus$ on the right. Also shown in the insert of each plot is the
  invariant mass below 3.0~\gevcc. As evident in the regions between the
  feed-down peaks and the $D_2^*$ signals, the fit quality is much improved by
  the addition of $S$-wave contributions for $D_0^*$.}
  \figlabel{d2_final}
\end{figure}

In \tabref{d2_summary} we compare our results for the narrow $D_2^*$ states and
our  results for the broad state, which are interpreted as $D_0^*$ states, with
the PDG~\cite{pdg:pdg2003} averages and with recent results from Belle.
\cite{Abe:2003zm}  In addition to the values in \tabref{d2_summary}, we also
measure the mass splitting between  $\dmeson_2^{*+}$ and $\dmeson_2^{*0}$ to be
$3.1\pm1.9\pm0.9$ compared to the PDG value of $0.0\pm3.3$. The $D_0^*$ results
from Belle only consider the neutral final state, so our  evidence for
$D_0^{*+}$ is the  first such observation. For both the $D_2^*$ and $D_0^*$ the
statistical accuracy of our results compare favorably with both the world
averages (if any) and the Belle results. The FOCUS results have recently been
published~\cite{Link:2003bd}, and the published paper should be consulted for
additional details.

\begin{table}
 \begin{center}
  \caption{Summary of results for $D_2^{*0}$ and $D_2^{*+}$. This table
compares the results from this measurement with those from the Particle Data
Group and Belle (not included in the PDG).}
  \tablabel{d2_summary}
  \begin{tabular}{|l|l|l|l|l|}\hline
    & 
    \multicolumn{1}{c|}{$\dmeson_2^{*0}$} &
    \multicolumn{1}{c|}{$\dmeson_2^{*+}$} &
    \multicolumn{1}{c|}{``$\dmeson_{0}^{*0}(j_\ell = \half)$''} &
    \multicolumn{1}{c|}{``$\dmeson_{0}^{*+}(j_\ell = \half)$''} \\\hline

    Yield & $5776 \pm 869 \pm 696$ & $3474 \pm 670 \pm 656$ 
     & $9810 \pm 2657$ & $18754\pm2189$ \\\hline
    Mass & $2464.5 \pm 1.1 \pm 1.9$ & $2467.6 \pm 1.5 \pm 0.8$  
     & $2407 \pm 21 \pm 35$ & $2403 \pm 14 \pm 35$ \\
    PDG03 & $2458.9 \pm 2.0$ & $2459 \pm 4$ & &\\
    Belle03 & $2461.6 \pm 3.9$ &  & $2308 \pm 36$ & \\\hline
    
    Width & $38.7 \pm 5.3 \pm 2.9$ & $34.1 \pm 6.5 \pm 4.2$    & $240 \pm 55 \pm 59$ & $283 \pm 24 \pm 34$ \\
    PDG03 & $23 \pm 5$ & $25^{+8}_{-7}$ &&\\
    Belle03 & $45.6 \pm 8.0$ & & $276 \pm 66$ & \\
    
    \hline	  
  \end{tabular}
 \end{center}
\end{table}

\section{Observation of $\dmeson_s^+(2317) \to \dsplus \pizero$} 

Recent observations by \babar,\cite{Aubert:2003fg} CLEO,\cite{Besson:2003cp}
and Belle~\cite{Krokovny:2003zq} of two unexpected, narrow, excited $D_s$
states have proved exciting. A likely explanation for these states appears to
be that they are $L=1$ mesons which lie below the $D^{(*)} K$ thresholds, the
preferred decay modes. Instead, they are constrained to decay via $D_s^{(*)+}
\pizero$. Why these $D_s$ states should be less massive than earlier predicted
is an open question. FOCUS confirms the first of these states, the
$D_s^+(2317)$ which decays to $\dsplus \pizero$. The \dsplus is reconstructed
in $\dsplus \to \phi \piplus$; $\phi \to \kplus \kminus$ which is then combined with a \pizero reconstructed from two
photons in the inner calorimeter.  The energy of this \pizero  is corrected
based on studies of the decays $D_s^{*+} \to D_s \pizero$ and $\dzero \to
\kminus \piplus \pizero$.  A plot of the $D_s \pizero$ mass is shown in
\figref{ds2317}. We find 58 events and preliminarily measure the mass to be
$2323\pm2$~\mevcc. We do not quote a systematic error at this time. 

\begin{figure}
  \begin{center}
  \includegraphics[width=16cm]{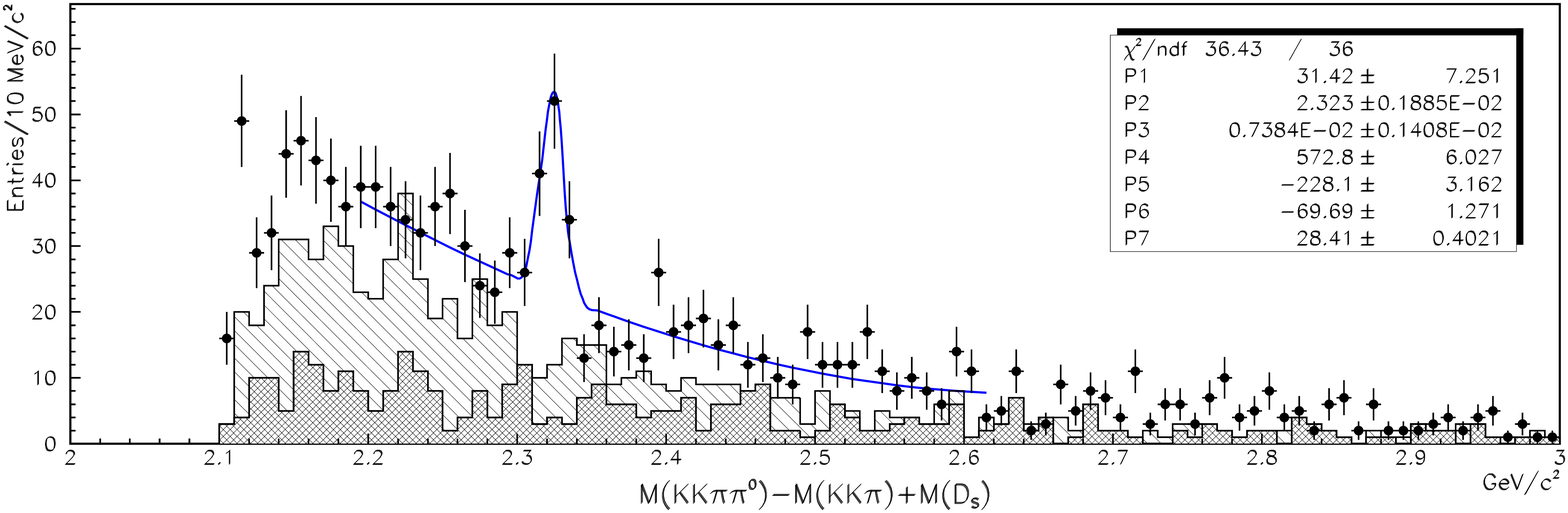}  
  \end{center}  
  \caption{Preliminary invariant mass plot showing $D_s^+(2317) \to D_s^+
  \pizero$. We observe 58 events in this decay mode with a mass of $2323\pm2$~\mevcc.}
  \figlabel{ds2317}
\end{figure}

\section{Conclusion}

Interest in charmed meson spectroscopy has increased recently because of
surprising discoveries in the $D_s$ sector. FOCUS is adding to the body of
knowledge for $L=1$ mesons with the measurements presented in this contribution. These
include precise measurements of the $D_2^*$ parameters with comparable
statistical precision to the previous world averages and the first evidence for the
expected broad $D_0^*$ states. Additionally we are able to confirm the
observation of the $D_s^+(2317)$ state observed by other experiments. Not presented in
this contribution are studies of $D_s^+(2536)$ and $D_s^+(2573)$ in $D^{(*)}K$
final states or studies of $L=1$ \dmeson mesons decaying to $\dsp \pion^\pm$.

\section*{Acknowledgments}
We wish to acknowledge the assistance of the staffs of Fermi National
Accelerator Laboratory, the INFN of Italy, and the physics departments of the
collaborating institutions. This research was supported in part by the U.~S.
National Science Foundation, the U.~S. Department of Energy, the Italian
Istituto Nazionale di Fisica Nucleare and Ministero della Istruzione
Universit\`a e Ricerca, the Brazilian Conselho Nacional de Desenvolvimento
Cient\'{\i}fico e Tecnol\'ogico, CONACyT-M\'exico, and the Korea Research
Foundation of the Korean Ministry of Education.

\section*{References}
\bibliographystyle{phaip}
\bibliography{physjabb,abbrev,pdg,focus,e687,theory,other,cleo}






\end{document}